# Observation of the Magnus Nonlinear Hall effect from Chiral Weyl Monopoles

Heda Zhang[1*], Nikolai Peshcherenko[1], Ning Mao[1], Nianlong Zou[2], Jiaqiang Yan[3], Claudia Felser[1*], and Yang Zhang[2,4*]

[1] Max Planck Institute for Chemical Physics of Solids, Dresden, Germany

[2] Department of Physics and Astronomy, University of Tennessee, Knoxville, TN 37996, USA

[3] Materials Science and Technology Division, Oak Ridge National Laboratory, Oak Ridge, TN, USA

[4] Department of Physics, National University of Singapore, 117551, Singapore

[*]Emails: Heda.Zhang@cpfs.mpg.de, Claudia.Felser@cpfs.mpg.de, yangzhang@utk.edu

**The nonlinear Hall effect (NLHE) connects crystalline symmetry to quantum geometry, offering a probe of band topology beyond linear transport. While most studies have focused on the Berry curvature dipole in low-symmetry crystals, mechanisms that directly probe Berry monopoles in higher-symmetry chiral lattices remain unexplored. Here, we report the observations of the NLHE in the chiral Weyl semimetal CoSi, a platform where the Berry curvature dipole is symmetry-forbidden. By employing focused ion beam–fabricated crossbar devices, we detect a robust second-harmonic Hall voltage under zero magnetic field, hosting all key signatures of the NLHE. Theoretical analysis attributes the nonlinear Hall conductivity to skew scattering of self-rotating electron wave packets, whose chirality is dictated by the underlying band topology-a process reminiscent of the classical Magnus effect. Furthermore, the NLHE signal exhibits a temperature-dependent sign reversal, and a strong, linearly field-dependent modulation that grows with carrier mobility, directly reflecting the topological Weyl nodes distribution near the Fermi level. These findings establish CoSi as a platform for Berry**

**monopole–driven nonlinear transport, demonstrating a skew-scattering route to topological nonlinear Hall responses that bypasses conventional symmetry constraints.**

Nonlinear transport has emerged as a critical frontier in the study of topological phases, enabling the direct interrogation of quantum geometry beyond the limits of linear response[1, 2, 3, 4, 5, 6, 7, 8, 9, 10, 11, 12]. Unlike the linear Hall effect, the NLHE arises solely from the breaking of inversion symmetry, making it relevant to a wide range of non-centrosymmetric materials[5]. Importantly, NLHE is closely linked to the electron wavefunction's geometrical properties – specifically the Berry curvature dipole[1, 2] and quantum metric[8, 11]. The paradigm in non-magnetic materials relies on the Berry curvature dipole (BCD), which has been successfully observed in $WTe_2$[3], $MoTe_2$[7] and elemental Tellurium[10]. In magnetic material $MnBi_2Te_4$, it has been shown that the magnetic field can switch both the net quantum metric dipole and its corresponding NLH conductivity[8]. Therefore, the NLH conductivity tensor (denoted by $\chi$) serves as a unique probe for exploring the quantum geometric properties of topological materials. Beyond these fundamental interests, the NLHE also offers significant potential for applications such as wireless, terahertz signal rectification and broadband frequency mixing[6, 10, 13, 14].

While the BCD approach has been fruitful, a critical challenge exists in materials with high crystalline symmetry, where the BCD is often constrained or forbidden. Chiral solids with non-trivial quantum geometry provide an ideal platform for studying the NLHE, since inversion symmetry is naturally broken[15, 16, 17, 18, 19, 20, 21, 22]. The concept of chirality in this class of materials can have significant consequences: it can lead to distinct chirality-dependent optical, mechanical, and electrical responses[7, 17, 19, 20, 23]. Notable examples include chiral fermions[16, 17], chiral phonons[24, 25, 26] and chirality-induced spin selectivity (CISS)[27, 28], which contribute to the broader, ongoing pursuit of understanding the connection between chirality and topology in solids. This is particularly relevant to Weyl physics, where the chiral symmetry is broken in two dual ways: through the separation of oppositely chiral Weyl nodes (Berry curvature monopoles) in reciprocal space, and through the breaking of mirror-symmetry of the crystal structure in real space. In this case, chiral crystal structure not only allows for chiral-dependent surface band selections, such as in CoSi and RhSi[16], but could also break the energy degeneracy of the Weyl node pairs, leading to circular photogalvanic effect[29, 30, 31].

Beyond these symmetry and topological considerations, the nonlinear response also brought renewed significance to the skew scattering mechanism by endowing it with an intrinsic character[32, 33, 34, 35, 36, 37]. This effect is aptly named the Magnus Hall effect, owing to its close analogy with the classical Magnus effect, in which the self-rotation of an object leads to an anomalous transverse velocity. In the linear regime, skew scattering depends solely on the asymmetry of the scattering potential and is therefore purely extrinsic[38]. In the nonlinear regime, the orbital angular momentum of a topological band induces a self-rotation of the electron wave packet, whose chirality is determined by the sign of its Chern number [Fig. 1(a)]. Consequently, the anomalous velocity arising from skew scattering, and hence the nonlinear Hall conductivity, is governed by the band topology. This is a quantum analog of the classical magnus effect[34]. In PT-symmetric antiferromagnets, anomalous skew scattering and Berry curvature can even cooperate to generate a Hall effect that is absent in the linear regime but emerges in the nonlinear regime[37]. These skew scattering mechanisms bypass the symmetry constraints inherent to the Berry curvature dipole and quantum metric[5], and directly probe the Berry curvature monopoles through nonlinear measurements.

In this work, we report the observation of nonlinear Hall effect in the chiral Weyl semimetal CoSi and establish it as an electronic version of the Magnus skew scattering effect[34]. Single crystal samples were carved into ~30×30 µm crossbar specimens [Fig. 1(c)] using focused ion beam (FIB) milling. We observed a clear second-harmonic Hall voltage response ($E^{2\omega}$), which has all key characteristics consistent with NLHE signals. Interestingly, we also found a systematic field-induced change for the second harmonic voltage response ($\Delta E^{2\omega}$), which has a linear dependence on the magnetic field, and becomes increasingly pronounced with rising carrier mobility. These experimental findings are interpreted and discussed within the framework of magnus skew scattering arising from two energetically separated Weyl monopoles close to the Fermi surface of CoSi. Our work demonstrates that the skew scattering mechanism can generate nonlinear Hall responses in high-symmetry chiral semimetals, and highlights this route to explore Berry curvature monopoles near the Fermi surface through nonlinear transport measurements.

The chiral crystal structure of CoSi supports a nonvanishing nonlinear Hall effect. As shown in Fig. 1(b), CoSi crystallizes in the chiral space group $P2_13$. The $2_1$-screw axis along the [100] direction (dashed line) generates zigzag chains, while the $C_3$ axis along the [111] direction enforces the overall three-fold rotation symmetry of the lattice. The combination of these symmetries forms the chiral cubic point group 23, which contains only proper rotations and lacks any inversion or mirror symmetries, thereby permitting a finite nonlinear Hall response. Crossbar-shaped regions of CoSi specimens were carved out using focused ion beam (FIB) for the transport measurements in this work[39]. Reducing the specimen's cross section – thereby increasing the current density – is crucial for observing clear NLHE in bulk, metallic samples. The flat surface of the specimen is the [111]-plane, as indicated by the triangular plateau of the crystal[39]. The $x'/y'$ axes of the crossbar form an angle of $\theta \approx 15°$ with respect to the [0$\bar{1}$1] crystal axis. In this experimental setup, the measured signal in the new coordinate system ($\chi_{y'y'x'}$) is equivalent to $\chi_{xyz}/\sqrt{3}$ in the standard Cartesian coordinate[39].

We first characterize the second harmonic voltage ($E^{2\omega}$) responses to an applied alternating current drive ($J^\omega$, $\omega$ = 71.717 Hz). Applying $J^\omega_{x'}$ between terminal 1 and 3 generates a transverse voltage ($E^{2\omega}_{y'}$) at double frequency with a 90-degree phase shift. As shown in Fig.

2(a), this voltage response exhibits a clear quadratic dependence on the drive current $J_{x'}^{\omega}$. Fig. 2(a) also presents the four-quadrant measurement that examines the polarity dependence between $E_{y'}^{2\omega}$ and $J_{x'}^{\omega}$. Comparing the first and second quadrants, we find that $E_{y'}^{2\omega}$ is independent of current directions. Comparing the first/second and fourth/third quadrant, we find that $E_{y'}^{2\omega}$ switches sign after swapping the voltage terminals. Lastly, we verify that the quadratic relation remains valid at $\omega$ = 171.717, 271.717 and 371.717 Hz, showing negligible frequency dependence [39]. Next, we followed by investigating the second harmonic voltage ($E^{2\omega}$) responses in different measurement geometries: a Hall-type geometry [Fig. 2(a)] and a resistance-type [Fig. 2(b)] geometry. Interestingly, the second harmonic voltage becomes vanishingly small when the same measurement is conducted in the resistance-type geometry. As shown in Fig. 2(b), with the drive current $J^{\omega}$ supplied between terminal 1 and 2, the second harmonic response $E^{2\omega}$ between terminals 3 and 4 is negligible. In other words, $E^{2\omega}$ depends critically on the Hall measurement geometry. These characteristics of the second harmonic Hall voltage response are consistent with those reported in previous NLHE studies[5].

The nonlinear Hall effect in non-centrosymmetric materials is expected to exhibit strong anisotropy, dictated by the underlying crystallographic symmetry of the material. For example, the NLH conductivity tensor ($\chi$) is maximized when the drive current is perpendicular to the Sb zigzag chains in BaMnSb$_2$[9], to the mirror line in WTe$_2$[3], and to the chiral axis in Tellurium[10]. Here, we examine the anisotropy of the second harmonic voltage response by swapping the voltage and current leads (i.e., $x' \leftrightarrow y'$). As shown in Fig. 2 (a) and (c), we find that the transverse second harmonic voltage response to equal current density is asymmetric with respect to $x' \leftrightarrow y'$. Since the crossbar device is four-fold symmetric, the anisotropy of the NLH conductivity is likely caused by further symmetry reduction by the crystalline lattice. We can extract the symmetric and the anti-symmetric components of the NLH conductivity tensor as $\chi^{sym/asym} = (\chi_{x'y'y'} \pm \chi_{y'x'x'})/2$. The symmetric component is larger in magnitude compared to the antisymmetric one, much like the behavior seen in another chiral material – elemental tellurium[39, 40].

We further investigate the temperature dependence of the anisotropic NLHE in CoSi. In Fig. 2(d), measured $J^\omega - E^{2\omega}$ relations at selected temperatures are plotted, revealing a clear difference in $J^\omega - E^{2\omega}$ relations depending on the current direction ($J_{x'}$ and $J_{y'}$). To quantify the difference, the measured $J^\omega - E^{2\omega}$ data are fitted to a second-degree polynomial [$E^{2\omega} = a + a_1 J^\omega + a_2 (J^\omega)^2$], and the quadratic coefficient ($a_2$) is converted to the nonlinear conductivity tensor by $\chi = \sigma^3 a_2 = \sigma^3 E^{2\omega}(J^\omega)^{-2}$. As shown in Fig. 2(e), for the two samples we have measured in this study, there is a clear sign reversal of $\chi$ occurring near T = 140 K. This sign reversal can be naturally accounted for by considering a pair of energetically separated (~11 meV) Weyl monopoles with opposite topological charge near the Fermi level. At low temperatures, the narrow energy window of integration encompasses only one of the two nodes, producing a positive NLH response; as the second node begins to contribute with increasing temperature, the net signal reverses. This interpretation is supported by our first-principles calculations and will be discussed in greater detail later.

It is also instructive to place the measured signal in the broader context of NLHE responses observed in other material systems. Nonlinear Hall responses have been reported across diverse material platforms, including thin films[41], van der Waals devices[3, 4, 6, 8, 42], and bulk crystals[12, 43, 44], but the reported values are often not expressed in the standard conductivity unit (A·V$^{-2}$), hindering direct comparison. In Fig. 3(c), we compile previously reported results using geometry-independent coordinates: the current density and the nonlinear Hall conductivity. Our measurements place CoSi among the strongest nonlinear Hall responses, in both bulk crystals (squares) and van der Waals devices (circles), achieved with modest current density. We find notable exceptions in moiré systems. For example, there are two consecutive works reporting a combined enhancement of nearly eight orders of magnitude in $\chi^{2D}$ [36, 45]. These results suggest that the nonlinear Hall effect in the correlated moiré systems may represent a qualitatively different regime.

Having established the temperature-dependent behavior of the nonlinear Hall response, we next turn to the effect of an external magnetic field. Although time-reversal (TR) symmetry breaking is not required for NLHE, the reciprocal question–namely, how does the NLH conductivity respond to an external magnetic field – remains largely unexplored. An exception is found in MnBi$_2$Te$_4$, in which the TR-symmetry is intrinsically broken by the Mn spins, and the NLH effect is shown to be TR-asymmetric[8]. Here, we also investigate the magnetic field dependence of $\chi$ in CoSi. We found that $E^{2\omega}$ is strongly dependent on the out-of-plane magnetic field strength ($B \perp [111]$) at low temperatures. In Fig. 3(a), we plot the field-induced change of the second harmonic voltage response as $\Delta E_{y\prime}^{2\omega} = E^{2\omega}(6.0\ T) - E^{2\omega}(0.0\ T)$ at selected temperatures. The field induced changes $\Delta E_{y\prime}^{2\omega}$ still follow a quadratic dependence on $J_{x\prime}^{\omega}$, and become vanishingly small near 70.0 K. The same measurement was performed at B = - 6.0 T, which causes the voltage change $\Delta E_{y\prime}^{2\omega}$ to switch sign (left panel). We perform the same polynomial fitting procedure to the field induced changes $\Delta E_{y\prime}^{2\omega} - J_{x\prime}^{\omega}$ and extract the change of nonlinear conductivity as $\Delta\chi$, which is shown in Fig. 3(b). The $\Delta E_{y\prime}^{2\omega}$ is clearly odd with respect to the magnetic field, and its sharp increase below 70 K coincides with a marked enhancement in electron mobility[46], suggesting a close link in their physical origin.

Interestingly, we found that the field-induced change $\Delta\chi$ is also anisotropic and break the $x'y'$-symmetry. This is shown in Fig. 3(c), where the right/left panels show $\Delta E_{y\prime}^{2\omega}$ and $\Delta E_{x\prime}^{2\omega}$ measured under the same field but with reversed drive-response directions ($x' \leftrightarrow y'$).

Compared to the zero-field data in Fig. 2(d), the asymmetry is even more pronounced. Following the previously described procedure to extract the anti-symmetric component of the field-induced voltage change, we find that the $\Delta E^{asym} - J^{\omega}$ relation also follows a clear quadratic dependence[39]. Using this relation, we then extract the field-induced nonlinear conductivity change tensor $\chi$ under B = 1.0 – 6.0 T and at temperatures of T = 10, 30, 60 and 100 K. The data presented in Fig. 3(d) clearly shows that $\Delta\chi$ follows a linear dependence on the magnetic field strength.

**Discussions**

As shown in Fig. 2(e), for both samples studied, a clear sign reversal occurs near T = 140 K. Such non-trivial temperature dependence strongly suggests that the observed nonlinear Hall effect should be related to intrinsic electronic band properties. While the Berry curvature dipole is the most established mechanism for NLHE, it is strictly forbidden in the cubic space group 198 (P2$_1$3) of CoSi due to the combination of cubic symmetry and topology of Fermi surface[47]. Moreover, the ratio of second harmonic to Drude current was observed (Fig. S3 (f)[39]) to be T-dependent even at low temperatures. This behavior contradicts the anticipated temperature-independent scaling at low temperatures, since both the Berry curvature dipole and Drude conductivity are proportional to the first power of relaxation time.

The temperature-induced sign reversal further rules out conventional skew scattering. Conventional skew scattering arises from non-centrosymmetric impurity potentials and typically maintains a constant sign unless the nature of the scattering center itself changes. For a sign reversal to occur via this trivial mechanism at high temperatures, one would require a highly improbable disorder configuration where impurities with opposite asymmetries are serendipitously positioned at different energy scales. Since this reversal is consistently observed across multiple samples with different disorder profiles, such an accidental explanation is unlikely.

Consequently, we attribute the observed response to the Magnus nonlinear Hall effect—a Weyl node chirality-dependent skew scattering mechanism. Crucially, while this effect is extrinsic in the sense that it requires scattering, the "handedness" of the scattering is dictated by the intrinsic topology of the Weyl nodes rather than the asymmetry of the impurity potential. This process serves as a quantum analog of the classical Magnus effect: the skew scattering arises from the self-rotation of the electron wave packet, a property directly fixed by the band's Chern number. As demonstrated by our model calculations (see Methods), inverting the Chern number inverts the sign of the Magnus Hall conductivity.

Our density functional theory (DFT) calculations [Fig. 4(a-c)] confirm that CoSi hosts two Weyl nodes close to Fermi level (denoted by $W_1$ and $W_2$), with opposite Chern numbers (+4 and -1, respectively), and are separated in energy by approximately 11 meV. The topological nature of these bands is further evidenced by the presence of surface bands as shown in Fig. 4(d). At low temperatures, NLH signal is dominated by the $W_1$ node. However, as the temperature rises, thermal broadening activates carriers from the $W_2$ node, which introduces a counter-propagating nonlinear Hall current. The energy separation between $W_1$ and $W_2$ (11 meV/ 127.6 K) is in a plausible agreement with the distance between two peaks in conductivity behavior at Fig. 2(e). For the nonlinear skew scattering contribution one can then estimate[34],

$$\sigma_{sk} \sim n e v_F \left(\frac{e\tau}{p_F}\right)^2 \frac{\tau}{\tilde{\tau}},$$

where $\tau, \tilde{\tau}$ describe momentum relaxation and skew scattering time correspondingly. Assuming realistic material parameters $n \sim 10^{20}\ cm^{-3}$, $v_F \sim 10^5\ m/s$, $\tau \sim 10^{-13}\ s$, the estimate gives $\sigma_{sk} \sim 10^{-2} \tau \tilde{\tau}^{-1} AV^{-2}$, providing a reasonable agreement with experimental data $\sigma_{exp} \sim 5 \times 10^{-3} AV^{-2}$.

Finally, regarding the magnetic field-dependent second harmonic measurement, we propose the following interpretation for the field induced changes in Fig. 3. When time-reversal symmetry is broken, a Hall response is generally expected in both the linear and nonlinear regimes. However, an important distinction exists between the two: since the second-order response scales quadratically with current, the polarity of the carriers no longer determines the sign of the effect. In other words, the linear Hall effect is dictated by the carrier charge, whereas the nonlinear Hall effect is governed by the topological charge of the bands. The field-induced change of the nonlinear conductivity can be viewed as an analog of the conventional Hall effect, except that it now probes a topological charge distribution near the Fermi level. The rapid increase in magnitude with rising carrier mobility further supports the skew-scattering origin of this response.

In summary, we have established the chiral Weyl semimetal CoSi as a robust platform for the Magnus nonlinear Hall effect—a nonlinear transport regime driven by the skew scattering of self-rotating electron wave packets. Our observations reveal a pronounced nonlinear Hall response characterized by quadratic current scaling, crystallographic anisotropy, and, most notably, a temperature-dependent sign reversal. We identify this reversal as the spectroscopic signature of the underlying topology, tracing the thermal competition between two energetically separated Weyl nodes with opposite chirality. The nonlinear response is robust under zero magnetic field and can be further modulated by an out-of-plane magnetic field, with the field-induced changes scaling linearly and correlating with carrier mobility. These observations are naturally explained by a skew-scattering mechanism of self-rotating electron wave packets, which directly probes the topological charge distribution near the Fermi level. Our work highlights CoSi as a model system for studying Berry monopole–driven nonlinear Hall transport, bypassing the symmetry constraints that limit the conventional Berry curvature dipole mechanism.

**Methods**

**DFT bands and Chern number calculation**

In this section, we employ density functional theory (DFT) calculations within the generalized gradient approximation (GGA) for the exchange–correlation potential, using the Perdew–Burke–Ernzerhof (PBE) parametrization[48]. The plane-wave energy cutoff is set to 500 eV, and the Brillouin zone is sampled using an 8×8×8 k-point mesh. Structural relaxations are performed until the residual forces on each atom are smaller than 0.01 eV/Å. Based on the converged DFT calculations, the ab initio Bloch wavefunctions are projected onto atomic-orbital-like Wannier functions using Wannier90[49]. The Wannier basis set includes the s, p, and d orbitals of Co atoms and the s and p orbitals of Si atoms. The resulting tight-binding Hamiltonian is then used to investigate the topological properties associated with Weyl points. The Chern numbers of the Weyl points are evaluated using the Wilson-loop method on a dense 100×100×100 k-point mesh to ensure numerical convergence.

Then, we demonstrate that the DFT bands of CoSi provide for at least 2 Weyl points being positioned close to the Fermi level. These points constitute the well-known multiple crossing point at $\Gamma$[50] and next to it at $W_2$ Weyl crossing (positioned at $(0.06367, 0.06332, 0.06429)$), see also Fig. 1(b) of the main text). The Wilson loop calculation results presented at Fig. 1(c) prove that the Chern number of $W_1$ and $W_2$ node are 4 and -1, respectively.

**Chirality mediated skew scattering description**

In this section we showcase how the sign of skew scattering Hall conductivity contribution is controlled by the corresponding electron's wave packet chirality. The asymmetric scattering probability $w_{k \to k'} = -w_{k \to k'}$ required for skew scattering is given by the formula[34]

$$w_{k \to k'} = \int \frac{d^3q}{(2\pi)^3} \text{Im}\{V_{kk'}V_{k'q}V_{qk}\}\delta(\varepsilon_k - \varepsilon_{k'})\delta(\varepsilon_{k'} - \varepsilon_q)$$

In the limit of isotropic scattering potential $V(r) = V_0 \delta(r - r_0)$ the expression for scattering probability simplifies:

$$w_{k \to k'} = V_0^3 \int \frac{d^3q}{(2\pi)^3} \text{Im}\{\langle\psi_k|\psi_{k'}\rangle\langle\psi_{k'}|\psi_q\rangle\langle\psi_q|\psi_k\rangle\}\delta(\varepsilon_k - \varepsilon_{k'})\delta(\varepsilon_{k'} - \varepsilon_q)$$

A simple topological band crossing could be described by two bands Hamiltonian:

$$H(k) = d_0(k)\sigma_0 + \boldsymbol{d}(k) \cdot \boldsymbol{\sigma}$$

so that the corresponding chirality term could be simplified to[51]:

$$\text{Im}\{\langle\psi_k|\psi_{k'}\rangle\langle\psi_{k'}|\psi_q\rangle\langle\psi_q|\psi_k\rangle\} = \frac{1}{4}\hat{\boldsymbol{d}}(k) \cdot (\hat{\boldsymbol{d}}(k') \times \hat{\boldsymbol{d}}(q))$$

where $\hat{\boldsymbol{d}}(k) = \boldsymbol{d}(k)/|\boldsymbol{d}(k)|$. Thus, inverting the topological chirality (formally equivalent to reversing the sign of the $\boldsymbol{d}(k)$ vector) leads to a sign inversion of the effective skew scattering probability. Consequently, the preferred direction of wave packet propagation is reversed. We identify this mechanism as the physical origin of the conductivity sign change observed at intermediate temperatures, arising from the thermal activation of Weyl nodes possessing opposite Chern numbers.

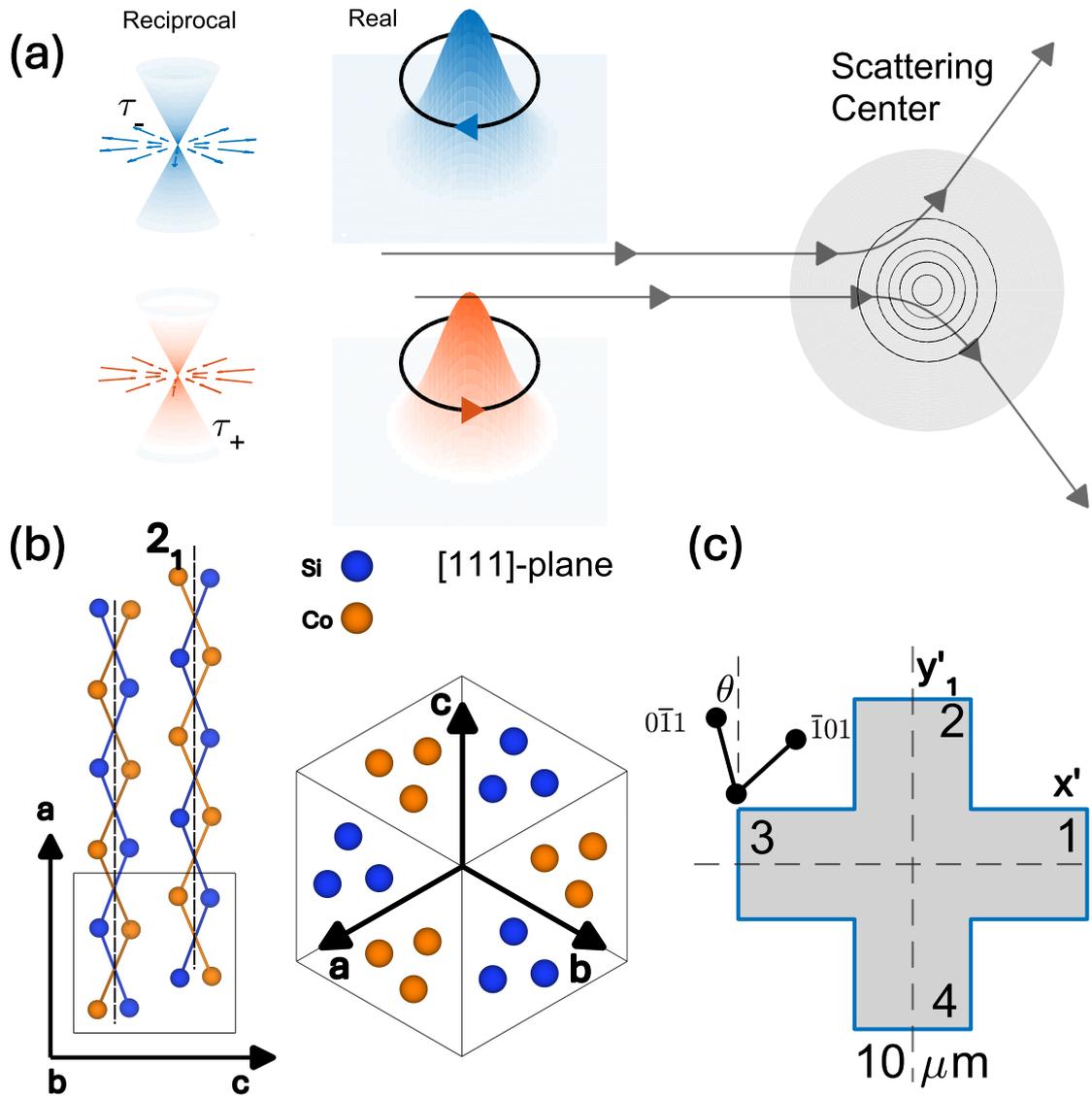

**Fig. 1 Illustration of the Magnus skew scattering nonlinear Hall effect, crystalline structure and device geometry of CoSi.** (a) The two Berry curvature monopoles with opposite chirality at separate energies endow the electron wave packet with opposite self-rotations. When propagating across a scattering center, electron wave-packets with opposite self-rotations (chirality) acquire opposite velocities due to chirality-mediated skew scattering, generating opposite contributions to the NLHE. (b) Crystalline structure of CoSi viewed along the [010] direction and [111] direction, illustrating the $2_1$ and $C_3$ symmetry. (c) Illustration of the crossbar device: the angle $\theta$ between channel bridges $x'$ and $[0\bar{1}1]$ is approximately 14.8 degrees, and $y'$ is orthogonal to $x'$.

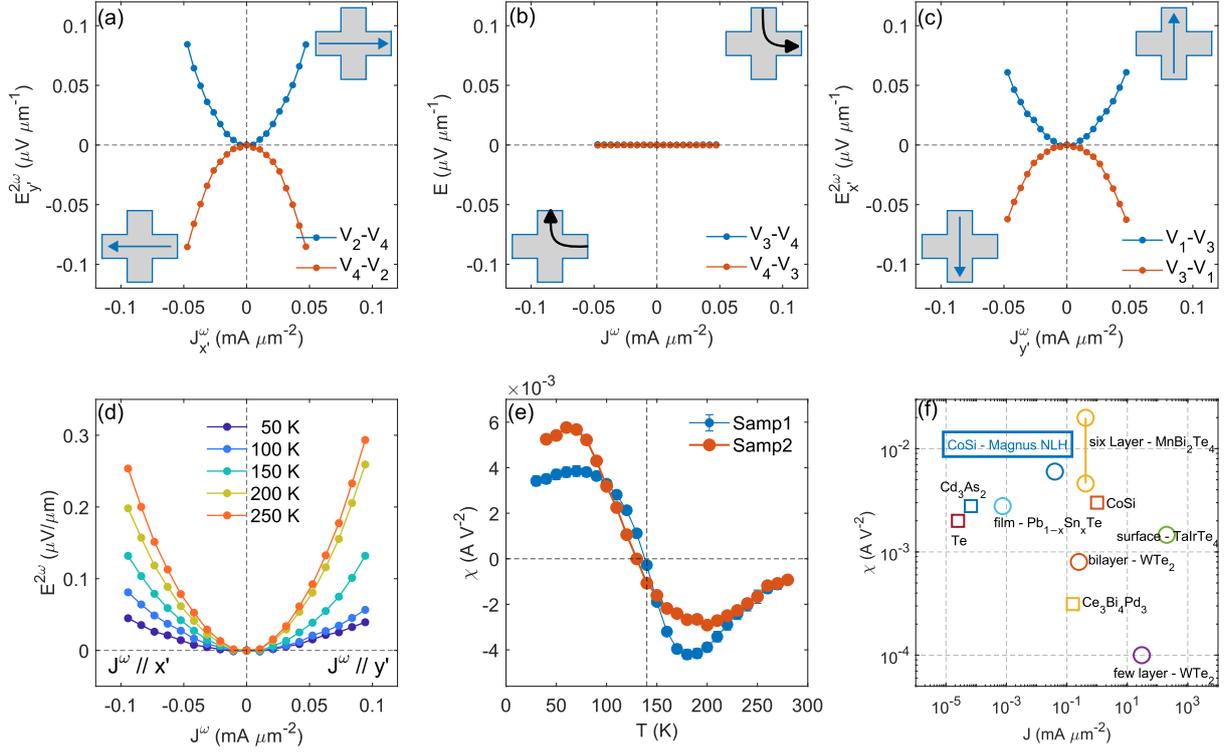

**Fig. 2 Key characteristics of the second harmonic voltage response.** (a-c) Second harmonic Hall and resistance voltage responses for currents applied along the $x'$ and $y'$ of the crossbar. The device shows a sizable voltage response in the (a, c) Hall measurement geometry, and negligible response in the (b) resistance measurement geometry. (d) Second harmonic Hall voltage-current ($E^{2\omega} - J_\omega$) relation showing clear anisotropic behavior in the full measurement temperature range. Only selected curves are shown for clarity. (e) The nonlinear Hall conductivity $\chi$ for the two samples measured in this study, showing a clear sign reversal around 140 K. (f) The measured $\chi$ of CoSi is compared with previously studied platforms to provide context and perspective.

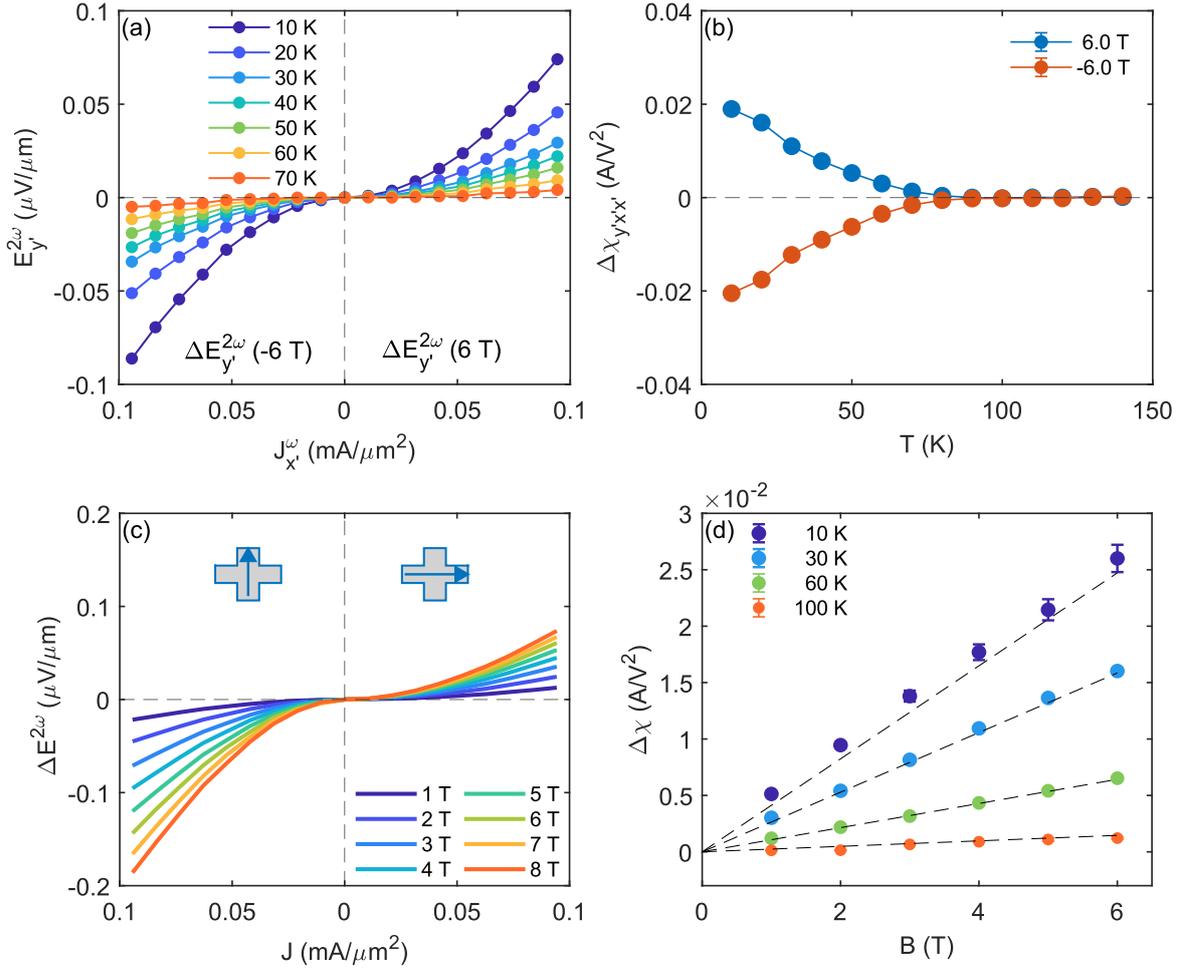

**Fig. 3 The magnetic field effect on the nonlinear Hall conductivity of CoSi.** (a) The out-of-plane field induced changes of second harmonic voltage response measured at B = ± 6.0 T for electric current along the $x'$ direction. (b) The extracted changes of nonlinear Hall conductivity ($\Delta\chi$) as functions of temperature at B = ± 6.0 T. (c) The field-induced second harmonic voltage change under different magnetic field strength, for electric currents applied parallel to the $x'$-direction (right) and $y'$-direction (left). (d) The field-dependence of the asymmetric nonlinear Hall conductivity change ($\Delta\chi$) at selected temperatures.

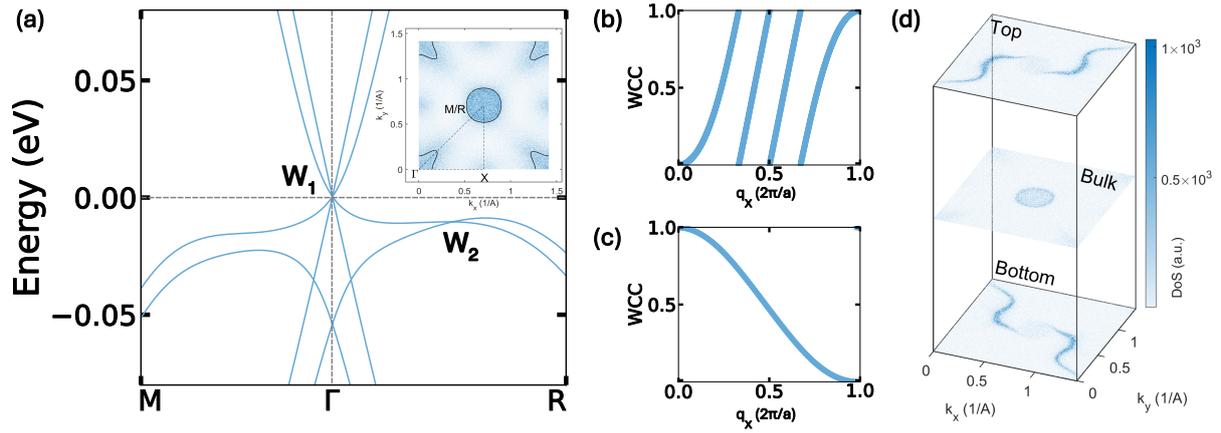

**Fig. 4 Density Functional Theory calculation of the band structure and topology of CoSi.**

(a) The band structure calculation of CoSi, shows two distinct Weyl points ($W_1$ and $W_2$) at $\Gamma$ and along the $\Gamma - R$ path. Inset shows the density of states at Fermi level, highlighting two main Fermi pockets near $\Gamma$ and $R$. (b, c) The calculated Wilson loop spectra for $W_1$ and $W_2$. The Wannier Charge Center (WCC) shows a crossing of four and one for $W_1$ and $W_2$, corresponding to Chern numbers of +4 and -1. (d) The topological nature of Weyl point $W_1$ is further corroborated by the surface band structure calculation, where clear Fermi arcs emerge after projecting onto the [001] surfaces.